\documentstyle[preprint, aps,epsf, floats,tighten,eqsecnum]{revtex}

\begin{document}

\preprint{
\vbox{\hbox{JHU--TIPAC--200001}
      \hbox{February 2000} }}
      
\title{Constraints on a Model with Pure \\
Right-Handed Third Generation Couplings}
\author{Adam Lewandowski}
\address{Department of Physics and Astronomy, The Johns Hopkins University \\
  3400 North Charles Street,  Baltimore, Maryland 21218}
\maketitle
\thispagestyle{empty}
\setcounter{page}{0}
\begin{abstract}%

We examine constraints on a model with pure right-handed third generation 
charged couplings.  The parameters of the right-handed mixing matrix and the 
right-handed coupling strength are constrained from semi-leptonic $B$ decays, 
the mass difference of neutral mesons, the CP violating observables 
$\epsilon$ and $\epsilon' / \epsilon$, and the electric dipole moment of the 
neutron.  We find the model to be tightly constrained by these parameters with 
several fine tuning conditions on the phases in the right-handed mixing 
matrix.  
There is also a necessarily non-zero value of the $W_{L}$-$W_{R}$ mixing 
parameter, $\zeta_{g}$.  CP 
asymmetry phases in neutral $B$ decays are discussed.

\end{abstract}

\newpage

\section{Introduction}

The standard $SU(2)_{L} \times U(1)$ model of the weak interactions has 
achieved 
great success.  Nevertheless, viable competing models with the $SU(2)_{L} 
\times 
SU(2)_{R} \times U(1)$ gauge group have been 
proposed~\cite{Mohapatra:rev,Langacker}.  In these models the 
left-handed Cabbibo-Kobayashi-Maskawa (CKM) mixing matrix is that of the 
standard model and the parameters of the right-handed mixing matrix as well as 
the right handed coupling and the mass of the right-handed gauge bosons are 
constrained by experiment.  In 1992, Gronau and Wakaizumi (GW) presented a 
model 
with this gauge group in which the flavor changing third generation decays 
occur 
only through right-handed currents~\cite{GW}.  This model owes its feasibility 
to the 
difficulty in differentiating between $(V-A)(V-A)$ quark-lepton couplings and 
$(V+A)(V+A)$ couplings.  Experimental evidence has since ruled out the GW 
model 
as a possible alternative to the standard model.  However, a more general 
choice 
than that chosen by GW for the right-handed mixing matrix, although tightly 
constrained by experiment, can not be entirely excluded on phenomenological 
grounds.

CP violation in the GW model and its more general extensions has been studied 
previously~\cite{GW,GronauCP,Hayashi,London:CPinBandep}.   Previous authors 
have 
shown that the GW 
model parameters are 
constrained by the CP violating observables $\epsilon$ and $\epsilon'$ and the 
neutron dipole moment.  They have also shown asymmetry values in nonleptonic 
neutral $B$ 
decays differing from standard model predictions.  It is, however, necessary 
to 
reexamine the constraints imposed by CP violation on these models in light of 
recent experiments~\cite{D0,CLEO}.

In this paper we take the following approach.  First, we briefly review
non-symmetric left-right models and the GW model.  Then, in section 3 we 
constrain the 
angles of the most general right-handed mixing matrix from 
observables not related to CP violation.  We find that there is a 
tightly constrained region in 
which this model is viable, and we make a particular choice of angles.  We 
then 
place constraints on the phases from CP violating observables
in section 4.  With constraints so imposed we examine various 
predictions in $B$ decays in section 5.  We 
summarize our results in section 6.

\section{Review of $SU(2)_{L} \times SU(2)_{R} \times U(1)$ models}

Langacker and Sankar have reviewed $SU(2)_{L} \times SU(2)_{R} \times U(1)$ 
models~\cite{Langacker}.  In discussing these models below we 
follow much of their notation.

In $SU(2)_{L} \times SU(2)_{R} \times U(1)$ models, the left and right-handed 
quarks and leptons transform under doublets of separate $SU(2)$ gauge groups.  
This gives rise to a covariant derivative of the form
\begin{equation} 
D^{\mu} = \partial^{\mu} + \frac{i}{2} ( g_{L} \tau^{a} 
W_{L}^{\mu a}+ 
g_{R} 
\tau^{a} W_{R}^{\mu a} + g^{\prime} Y B^{\mu}),
\end{equation}
where $g^{\prime}$ is the $U(1)$ gauge coupling, $\tau^{a}$ are 
the Pauli spin 
matrices, $W_{L, R}^{a}$ and $B$ are the gauge boson fields and $g_{L,R}$ are 
the $SU(2)_{L}$ and $SU(2)_{R}$ gauge coupling constants.  The gauge symmetry 
is 
spontaneously 
broken by introducing a Yukawa interaction with some Higgs sector and giving 
the 
Higgs a vacuum expectation value.  This gives masses to the quarks, leptons 
and 
gauge bosons.  We take a Higgs, $\Phi$, that transforms as $\Phi \to L \Phi 
R^{\dagger}$ under $SU(2)_{L}$ and $SU(2)_{R}$ and is neutral under 
hypercharge. 
 A general choice for the vacuum expectation value gives
\begin{equation}
 \Phi = \left( \begin{array}{ll}
\phi_{1}^{0}  &  \phi_{1}^{+} \\		
\phi_{2}^{-}  &  \phi_{2}^{0} 
         \end{array} \right)
      \to 
\left( \begin{array}{ll}
k & 0 \\
0 & k^{\prime}  
      \end{array} \right).
 \end{equation}  
With this Higgs we have the relation $M_{R} =  g_{R}/g_{L} M_{L}$, where 
$M_{L}$ and $M_{R}$ are the masses of the left and right-handed charged gauge 
bosons respectively.  Taking the ratio $g_{R}/g_{L}$ to be ${\cal{O}}(1)$, it 
is 
necessary to introduce additional Higgs to arrange for these masses to be much 
different.  Minimally, one introduces two 
doublets or triplets under $SU(2)_{L}$ and $SU(2)_{R}$ which carry a 
hypercharge 
of 1.  These obtain vacuum expectation values $v_{L}$ and $v_{R}$.

The Yukawa couplings to the quarks are given as
\begin{equation}
 - {\mathcal{L}}_{Y} = \sum_{i,j} \left(
\bar{f}^{\prime}_{iL} (r_{ij} 
\Phi 
+ s_{ij} \tilde{ \Phi} ) f^{\prime}_{jR} + {\rm h.c.} \right),
\end{equation}
where $f^{\prime}$ are the gauge eigenstate quark fields, $r$ and $s$ are 
general 
complex matrices and $\tilde{ \Phi} = \tau^{2} \Phi^{*} \tau^{2}$.  This term 
gives rise to the mass matrices, $M^{u} = rk+sk^{\prime*}$ and $M^{d} = r 
k^{\prime} + s k^{*}$.
In the mass basis of quarks and leptons the charged current 
interaction is 
given by
\begin{equation}
-{\mathcal{L}}_{CC} = \frac{g_{L}}{\sqrt{2}} \bar{u}_{iL} 
\gamma_{\mu} 
V^{L}_{ij} d_{jL} W^{\mu +}_{L} + \frac{g_{R}}{\sqrt{2}} 
\bar{u}_{iL} 
\gamma_{\mu} V^{R}_{ij} d_{jL} W^{\mu +}_{R} + {\rm h.c.} ,
\end{equation}
where $V^{L}$ and $V^{R}$ are the unitary mixing matrices for 
the 
quarks, the elements of which are
\begin{equation}
V^{L,R} = \left( \begin{array}{ccc} V_{ud}^{L,R} & V_{us}^{L,R} & V_{ub}^{L,R} 
\\
				 V_{cd}^{L,R} & V_{cs}^{L,R} & V_{cb}^{L,R} \\
				  V_{td}^{L,R} & V_{ts}^{L,R} & V_{tb}^{L,R} 
\end{array} \right).
\end{equation}

The kinetic term for the Higgs gives a mass structure to the 
gauge bosons.  There are two heavy neutral gauge bosons, the 
massless photon and 
charged gauged bosons from the left and right-handed sectors.  
The non-diagonal 
mass matrix of the charged left and right gauge bosons is 
\begin{equation}~\label{massmatrix}
M_{W}^{2} = \left( \begin{array}{cc}
	\frac{1}{2} g_{L}^{2}( |v_{L}|^{2} + |k|^{2} + 
|k^{\prime}|^{2}) & 
-g_{L} g_{R} k^{\prime} k^{*} \\
	-g_{L} g_{R} k^{\prime *} k &  \frac{1}{2} 
g_{R}^{2}( 
|v_{R}|^{2} + |k|^{2} + |k^{\prime}|^{2})
	\end{array} \right).
	\end{equation}
where $M_{L}$ and $M_{R}$ are the upper and lower diagonal elements 
respectively.  This matrix gives the mixing between the mass and the gauge 
eigenstates.  In terms of the mixing angle this is
\begin{equation}
\left( \begin{array}{c} W^{+}_{L} \\ W^{+}_{R} \end{array} 
\right) =
\left( \begin{array}{cc} \cos \zeta & - \sin \zeta \\
		    e^{i \omega} \sin \zeta & e^{i \omega} \cos 
\zeta
\end{array} \right)
\left( \begin{array}{c} W^{+}_{1} \\ W^{+}_{2} \end{array} 
\right),
\end{equation}
with
\begin{equation}
\tan 2 \zeta = \frac{2 g_{L} g_{R} |k^{\prime} k| }{M_{R}^{2}-M_{L}^{2}}.
\end{equation}

In this paper we will often use the following quantities:
\begin{equation}
 \zeta_{g} \equiv \frac{g_{R}}{g_{L}} \zeta,
\qquad \beta_{g} \equiv \frac{g_{R}^{2}}{g_{L}^{2}} \beta =  
\frac{g_{R}^{2} 
M_{1}^{2}}{g_{L}^{2} M_{2}^{2}}.  
\end{equation}
where $M_{1}$ and $M_{2}$ are the eigenvalues of the mass matrix 
(\ref{massmatrix}). In the case where $M_{R} \gg M_{L}$ one has $M_{1} \approx 
M_{L}$ and $M_{2} 
\approx M_{R}$.    $\zeta_{g}$ is the mixing parameter which determines the 
strength of the interactions due to mixing between left-handed and 
right-handed 
currents relative to pure left-handed current interactions.  
$\beta_{g}$ determines the relative strength of right-handed to left-handed 
interactions.

Gronau and Wakaizumi proposed to modify the mixing matrices such that the 
third 
generation of quarks couples to the other generations only through the 
right-handed $W$ bosons~\cite{GW}.  The specific parametrization is
\begin{equation}
V^{L} = \left( \begin{array}{ccc}~\label{leftCKM}
\cos \theta_{c}  & \sin \theta_{c} & 0 \\
-\sin \theta_{c} & \cos \theta_{c} & 0 \\
 0               &         0       & 1 \end{array} \right),
\end{equation}
for the left-handed mixing matrix, where $\theta_{c}$ is the 
Cabibbo angle.    In the original GW model the right-handed mixing matrix was 
parametrized by a single angle and CP violation accommodated by a single 
phase.  We choose to study the most general form of the right-handed mixing 
matrix, of which the GW model is a particular choice of angles and phases.  
This matrix involves three angles and four phases:
\begin{equation}
V^{R} = \left( \begin{array}{ccc}~\label{rightCKM}
c_{12} c_{13} e^{i(\alpha+\beta)} & - c_{13} s_{12} e^{i(\gamma+\alpha)} & 
s_{13} e^{-i (\beta + \gamma - \alpha)} \\
(-c_{12} s_{23} s_{13} + s_{12} c_{23} e^{i \delta}) e^{i (\beta-\alpha)} &
(s_{12} s_{23} s_{13} + c_{12} c_{23} e^{i \delta}) e^{i (\gamma - \alpha)} &
s_{23} c_{13} e^{-i (\beta+\gamma+\alpha)} \\
(-c_{12} c_{23} s_{13} - s_{12} s_{23} e^{i \delta}) e^{i \beta} &
(s_{12} c_{23} s_{13} - c_{12} s_{23} e^{i \delta}) e^{i \gamma} &
c_{23} c_{13} e^{-i (\beta+ \gamma)}
\end{array} \right),
\end{equation}
where $c_{ij}$ denotes $\cos \theta_{ij}$ and $s_{ij}$ denotes $\sin 
\theta_{ij}$.
We will denote the model employing (\ref{leftCKM}) and (\ref{rightCKM}) as the 
generalized GW model.  In the following section we will choose the angles in 
this 
matrix to satisfy experimental constraints.

\section{General Constraints on the Model}

The $b$ quark decays through pure right-handed couplings in this model.  This 
allows us to use constraints from semi-leptonic $B$ decays.  In particular 
the decay $b \to c l \nu$  gives the important relation
\begin{equation} \label{constraint} 
|V_{cb}^{R}|(\beta_{g}^{2} + \zeta_{g}^{2})^{1/2} = 
|V_{cb}^{SM}|= 0.036-0.042,
\end{equation}
where SM denotes the standard model value.  Results from CLEO measure the 
asymmetry in the decay $B \to D^{*} l \nu$ assuming a pure left-handed lepton 
current~\cite{CLEO}.  This puts an upper bound on the ratio
\begin{equation} \label{CLEOratio}
 \left( \frac{\zeta_{g}}{\beta_{g}} \right)^{2} < 0.30.  
\end{equation}
D0 performed direct searches for $W^{R}$~\cite{D0}.  For $g_{R} = g_{L}$ they 
obtain 
$M_{W_{R}} > 720$ GeV, which in turn implies $\beta_{g} < 0.012$.  Because 
$|V_{cb}^{R}| < 1$ by unitarity, condition (\ref{constraint}) gives $\beta_{g} 
> 
0.03$.  It is obvious that these two conditions can not simultaneously apply.  
It has been noted that the form of the right-handed mixing matrix and the size 
of the ratio $g_{R}/g_{L}$ affect the lower bound on the mass of 
$W_{R}$~\cite{Rizzo}.  In 
particular for $|V_{ud}^{R}| \ll 1$ and $g_{R} > g_{L}$ one lowers the bound 
on 
the $W_{R}$ mass.  In the region of parameter space of $V^{R}$ in which 
$|V_{ud}^{R}| \ll 1$ we apply the constraints to be discussed below and find 
that the neutral $B_{s}$ mass difference is too small to satisfy the 
experimental lower bound.  We are left with the 
region where $g_{R} > g_{L}$, which may be unnatural in grand unified 
models~\cite{Rizzo}.  D0 presents constraints on $M_{R}$ for $g_{R}/g_{L} = 
\sqrt{2}$ but for no higher values of the ratio.  It is not unreasonable to 
accept of $\beta_{g}$ in the range $0.03-0.04$ for
$g_{R}/g_{L} \sim 2$.  In this paper we choose $\beta_{g} = 0.035$.  
Additionally, there is the ratio from semi-leptonic $B$ decays
\begin{equation}
\left| \frac{V_{ub}^{R}}{V_{cb}^{R}} \right| 
= \left| \frac{V_{ub}^{SM}}{V_{cb}^{SM}} \right| = 0.09 \pm 0.03.
\end{equation}
This gives the constraint $s_{13}/s_{23}c_{13} = 0.09 \pm 
0.03$.

We now turn to the constraints imposed by the mass difference of the neutral 
mesons $K$, $B_{d}$ and $B_{s}$.  The neutral $K$ mass difference is given as
\begin{equation} \label{mk0}
\Delta m = -2 {\cal{R}}e \langle K^{0}|H_{\Delta S=2} | \bar{K^{0}} \rangle .
\end{equation}
The effective $\Delta S = 2$ Hamiltonian arises through the box diagram.  In 
this model there are contributions due to the exchange of two $W_{R}$'s and 
the 
exchange of a $W_{R}$ and a $W_{L}$ in addition to the two $W_{L}$ 
exchange familiar from the standard model.  We use the result of Mohapatra 
{\em 
et 
al.} for the box diagram without QCD 
corrections~\cite{boxdiagram}.  
To calculate the matrix 
element of quark field operators we use the vacuum insertion approximation 
with bag factors equal to unity. The neutral $B_{d}$ and $B_{s}$ mass 
differences are determined by 
relations similar to (\ref{mk0}) and we again use the results 
\cite{boxdiagram}. 
 We 
use estimates of the $B_{d}$ and $B_{s}$ decay constants from the 
lattice~\cite{lattice}.  For 
the experimental values
\begin{eqnarray}
\Delta m_{K} & = & 3.49 \times 10^{-12} \, {\rm MeV}, \\
\Delta m_{B_{d}} & = & 3.05 \times 10^{-10} \, {\rm MeV}, \\
\Delta m_{B_{s}} & > & 8.16 \times 10^{-9} \, {\rm MeV},
\end{eqnarray}
we find the following satisfactory choice of 
angles:
\begin{equation}
\theta_{13} = 0.08, \qquad
\theta_{12} = -0.04, \qquad
\theta_{23} = 1.8,
\end{equation}
 where tuning of ${\cal{O}}(10^{-2})$ is necessary for $\theta_{13}$ and 
$\theta_{12}$.  Tuning of ${\cal{O}}(10^{-1})$ is necessary for $\theta_{23}$.  
With these angles we 
satisfy the experimental conditions within theoretical 
uncertainties.  We will use these values in the remainder of this paper.  At 
this point there 
are no constraints on the four phases in this model.  These will be adjusted 
by the CP violating observables discussed in the next section.

\section{Constraints from CP Violation}~\label{CPconstraint}

CP violation has been measured in the neutral $K$ sector in the form of the 
observables $\epsilon$ and $\epsilon'$.  In addition, CP violation should give 
rise to a nonzero electric dipole moment of the neutron.  We will now use the 
measurements of $\epsilon$, $\epsilon' / \epsilon$ and the upper bound on the 
neutron dipole moment to constrain the phases in our model.

The parameter $\epsilon$ is related to the $\Delta S = 2$ Hamiltonian.
\begin{equation}
\epsilon = \frac{e^{i \pi/4}}{\sqrt{2}} \frac{{\cal{I}}m \langle 
K^{0}|H_{\Delta 
S=2} | \bar{K^{0}} \rangle}{\Delta m_{K}}.
\end{equation}
The effective Hamiltonian is that used to calculate the neutral $K$ mass 
splitting.  Using this and  
the choice of angles from the previous section we find several terms of 
${\cal{O}}(10^{-1})$.  The dominant contribution is 
from the box diagram 
involving the exchange of two $W_{R}$ and two top quarks.  This gives a strong 
contribution in this model because $V_{td}^{R} \sim 0.23$ and $V_{ts}^{R} \sim 
1$, whereas in the 
standard model the two top exchange diagram is CKM suppressed.  The leading 
terms in $\epsilon$ are
\begin{equation}
| \epsilon | = \left| 0.14 \cos 2(\beta-\gamma) \sin \delta - 0.14 \sin 
2(\beta-\gamma) 
\cos \delta +0.23 \sin (\beta-\gamma-\delta) \right|.
\end{equation}
With the experimental value of $(\epsilon = 2.28 \pm 0.02) \times 10^{-3}$  
and assuming no cancellation between terms, this suggests
\begin{equation}~\label{phase1}
\sin (\beta - \gamma)  \alt   {\cal{O}}(10^{-2}), \qquad 
\sin \delta  \alt  {\cal{O}}(10^{-2}).
\end{equation}
There are other terms in the calculation of $\epsilon$ of order less than or 
equal to $10^{-2}$ which impose no further constraints on the phases.

The parameter $\epsilon^{\prime}$ is given in terms of the $K$ 
decay 
amplitude to two pions as
\begin{equation}~\label{analeprime}
\epsilon^{\prime} = \frac{e^{i(\frac{\pi}{2} + \delta_{2} - 
\delta_{0})}}{\sqrt{2}} \frac{{\cal{R}}e A_{2}}{{\cal{R}}e 
A_{0}} \left( 
\frac{{\cal{I}}m A_{2}}{{\cal{R}}e A_{2}} - \frac{{\cal{I}}m 
A_{0}}{{\cal{R}}e 
A_{0}} \right),
\end{equation}
with
\begin{equation}~\label{isoamp}
 A_{i} = \langle (\pi \pi)_{I=i} | H_{\Delta S = 1} | K^{0} \rangle,
\end{equation}
where $i$ denotes the isospin channel and $\delta_{i}$ is the 
hadronic phase 
shift.  The problem of calculating $\epsilon^{\prime}$ is then 
to calculate 
the $\Delta S = 1$ Hamiltonian and with this, to estimate the 
decay 
amplitudes.  Of course, this problem is plagued with hadronic 
uncertainties.  The calculation of $\epsilon'$ in this model is interesting 
but lengthy.  We relegate it to the appendix.

We find in the resulting expression for $\epsilon' / \epsilon$ terms 
proportional to 
$\beta_{g}$ and $\zeta_{g}$.  The terms proportional to $\beta_{g}$ are too 
small 
to accommodate the measured value of $\epsilon' / \epsilon$ of $(21.2 \pm 4.6) 
\times 10^{-4}$.  Terms proportional to $\zeta_{g}$ must provide the dominant 
contribution.  We will see the effects of a non-zero $\zeta_{g}$ in the 
following section.  The 
dominant terms are
\begin{equation} \label{epsprime}
\left| \epsilon^{\prime}/ \epsilon \right|= \left| \zeta_{g} \left( 1.8 
R_{c}^{LR} \sin 
(\alpha-\beta) 
-4.0 R_{u}^{LR} \sin(\alpha + \beta) \right) \right|,
\end{equation}
where $R_{u}^{LR}$ and $R_{c}^{LR}$ are ratios of operators defined in the 
appendix. They are estimated to be ${\cal{O}}(1)$ and ${\cal{O}}(10^{-1})$ 
respectively.  The constraint from $\epsilon' / \epsilon$ then requires either 
small phases or a small $\zeta_{g}$.

The electric dipole moment of the neutron arises in this model at the one loop 
level due to mixing of the $W_{L}$ and $W_{R}$.  In the standard model one 
loop 
diagrams do not contribute because they are proportional to the 
magnitude of CKM elements and so are real.    $W_{L}$-$W_{R}$ 
mixing permits imaginary coefficients in the loop diagram, allowing for a 
non-zero 
edm.  Electric dipole moments of the $u$ and $d$ quark arise from diagrams 
involving the creation of a virtual $W$ and the emission of a photon.  These 
contributions have been calculated \cite{edm} and are given as
\begin{eqnarray}
d_{u} & = & \frac{e G_{F}}{4 \pi^{2}} \zeta_{g} \sum_{j=d,s,b} m_{j} 
{\cal{I}}m 
(V_{uj}^{L} V_{uj}^{R*}) f_{1} \left( \frac{m_{j}^{2}}{M_{L}^{2}} \right), \\
d_{d} & = & \frac{e G_{F}}{4 \pi^{2}} \zeta_{g} \sum_{j=u,c,t} m_{j} 
{\cal{I}}m 
(V_{jd}^{L} V_{jd}^{R*}) f_{2} \left( \frac{m_{j}^{2}}{M_{L}^{2}} \right),
\end{eqnarray}
where $f_{1}$ and $f_{2}$ are dimensionless functions of the quark masses.  In 
addition to the loop diagram there is also a contribution from the exchange of 
a 
mixed $W_{L}$-$W_{R}$ from the $u$ to the $d$ with the emission of a photon.  
This contribution has large hadronic uncertainties and is estimated in the 
harmonic 
oscillator parton model of the neutron~\cite{edm,oscillator}.  It is given by
\begin{equation}
d_{ex} = \frac{e G_{F}}{3 \pi^{3/2}} \zeta_{g} \sqrt{2 m_{q} \omega} 
(1-\beta_{g}) {\cal{I}}m (V^{L}_{ud} V_{ud}^{R*}),
\end{equation}  
where we use $\sqrt{m_{q} \omega} = 0.3$ GeV.  The edm of the neutron is 
related 
to these contributions by
\begin{equation}
d_{n} = \frac{4}{3} d_{d} - \frac{1}{3} d_{u} + d_{ex},
\end{equation}
Evaluated with our choice of angles we find
\begin{equation}
d_{n} = \zeta_{g} \left( -3.9 \times 10^{-21} \sin (\alpha+\beta) - 2.3 \times 
10^{-22} 
\sin (\alpha + \gamma) \right) \hbox{{\rm e-cm}}
\end{equation}
The experimental upper bound is $d_{n} < 2.6 \times 10^{-25}$ e-cm.  This 
constraint could be accommodated by a small $\zeta_{g} \alt 
{\cal{O}}(10^{-5})$. 
However, this would make $\epsilon' / \epsilon$ too small.  Together the edm 
and 
$\epsilon' / \epsilon$ require
\begin{equation}
\zeta_{g} \sim {\cal{O}}(10^{-2}), \qquad
\sin (\alpha+\beta ) \sim {\cal{O}}(10^{-3}).
\end{equation}
We see that $\zeta_{g}$ must be close to the upper bound of (\ref{CLEOratio}).

\section{Discussion on $B$ decays}

We have now found a satisfactory albeit tightly constrained region in which 
the generalized GW
model is valid.  If we make the simplifying assumption
\begin{equation}~\label{equality}
\beta = \gamma = -\alpha, \qquad \delta=\pi,
\end{equation}
which is consistent with the constraints and necessarily correct to at least 
${\cal{O}}(10^{-2})$, the right-handed mixing matrix becomes
\begin{equation}~\label{finalV}
V^{R} = \left( \begin{array}{ccc} 
   0.996 & 0.0399    & 0.0799 e^{-i 3 \beta} \\
-0.0862 e^{i 2 \beta} & -0.195 e^{i 2 \beta} & 0.977 e^{-i \beta} \\
 -0.233 e^{i \beta}  & 0.979 e^{i \beta}   & -0.198 e^{-i 2 \beta}
\end{array} \right).
\end{equation}
Although we found no others in our search, we do not suggest that this is the 
only possible choice of the seven 
free parameters in this matrix that satisfy the experimental constraints.  We 
merely point out 
that this particular choice is phenomenologically acceptable and as such the 
GW 
ansatz of pure right-handed $b$ decays is not completely ruled out.  We now 
examine some consequences of this choice of parameters.

The ratio of branching ratios
\begin{equation}~\label{branchingratio}
R=\frac{{\rm Br} (B^{-} \to \psi \pi^{-})}{{\rm Br} (B^{-} \to \psi K^{-})} = 
0.052 \pm 0.024.
\end{equation}
has been measured.  In the limit of dominant right-handed tree contributions 
to 
the decay we have
\begin{equation}
R \approx \left| \frac{V_{cd}^{R}}{V_{cs}^{R}} \right|^{2} = 0.2,
\end{equation}
where the ratio has been evaluated according to (\ref{finalV}).  To what 
degree 
should we trust the discrepancy here between experiment and our model?  
Certainly the ratios are the same within an order of magnitude.  Penguin 
contributions 
will affect the theoretical prediction.  There will also be a strong 
contribution due to $W_{L}$-$W_{R}$ mixing because of the relatively large 
value of $\zeta_{g}$. 
 To estimate the mixing contribution in the decay $B^{-} \to \psi 
K^{-}$ we look at the ratio of mixed to unmixed tree level contributions
\begin{equation}
\frac{\zeta_{g}}{\beta_{g}} \times \frac{V_{cs}^{L}}{V_{cs}^{R}} \times 
\frac{\langle O^{LR} \rangle }{\langle O^{RR} \rangle} \alt 3.5,
\end{equation}
where we have used the upper bound (\ref{CLEOratio}) and set the ratio of 
matrix 
elements \begin{equation}
\frac{\langle O^{LR} \rangle} {\langle O^{RR} \rangle} =
\frac{\langle K^{-} \psi | \bar{b} \gamma^{\mu} 
(1+\gamma^{5}) c \bar{c} \gamma_{\mu} (1-\gamma^{5} ) s | B^{-} \rangle 
}{\langle K^{-} \psi | \bar{b} \gamma^{\mu} 
(1+\gamma^{5}) c \bar{c} \gamma_{\mu} (1+\gamma^{5} ) s | B^{-} \rangle }=1.4,
\end{equation}
 found in the 
vacuum insertion approximation.  There is a 
substantial and possibly dominant contribution to the decay due to mixing.  
Our choice of phases and angles is then consistent with the ratio 
(\ref{branchingratio}).   

In the neutral $B$ meson system one can write the physical mass 
eigenstates in 
terms of the gauge eigenstates as~\cite{Nir:Flavor}
\begin{equation} |B^{0}_{1,2} \rangle = p |B^{0} \rangle \pm q | \bar{B^{0}} 
\rangle.
\end{equation}
Their decay amplitudes are
\begin{eqnarray}
A & = & \langle f | H | B^{0} \rangle \qquad {\rm and} \\
\bar{A} & = & \langle f | H | \bar{B^{0}} \rangle,
\end{eqnarray}
where $f$ is a CP eigenstate.  If there is a single dominant 
decay process 
(e.g. no strong penguin processes), then the decay asymmetry 
becomes~\cite{Gronau:asymm}
\begin{equation}
a_{f} = \frac{ \Gamma(B_{phys}^{0} \to f) - 
\Gamma(\bar{B}_{phys}^{0} \to f) 
}{\Gamma(B_{phys}^{0} \to f) + \Gamma(\bar{B}_{phys}^{0} \to
f)}
\propto {\cal I }m \left( \frac{q \bar{A}}{p A} \right),
\end{equation}
where $B_{phys}^{0} (\bar{B}_{phys}^{0})$ denotes the time 
evolved $B^{0} (\bar{B^{0}})$ meson.
If the final state, $f$, is not a CP eigenstate but a neutral 
meson such as $K_{S}$, then it 
also contributes a mixing term to the asymmetry.

In the absence of mixing $(\zeta_{g}=0)$ all $B$ decays occur 
through pure right-handed interactions.  There are both tree and penguin 
diagrams contributing to the decay.  However,  due to the phase constraints 
imposed on this model, tree diagrams and penguin diagrams contribute with the 
same phase up to corrections to which (\ref{equality}) holds.  This allows for 
clean asymmetry predictions which have been previously 
discussed~\cite{GronauCP,Hayashi}.

However, in a situation in which mixing is large it is necessary to consider 
pollution 
from mixing.  Previous work is no longer applicable to this situation.  In the 
case of 
$B_{d}^{0} \to 
\psi K_{S}$, we see that the ratio of left-right to
right-right tree amplitudes is given by
\begin{equation}
\left| \frac{T^{LR}}{T^{RR}} \right| = \left| \frac{V_{cb}^{R} 
V_{cs}^{L*}}{V_{cb}^{R} V_{cs}^{R*}} \right | \times 
\left| \frac{\zeta_{g} }{\beta_{g}} \right| \times \left| 
\frac{ \langle O^{LR} \rangle }{ \langle O^{RR} \rangle } \right|,
\end{equation}
where
\begin{eqnarray}
\langle O^{RR} \rangle & = & \langle K_{S} \psi | \bar{b} \gamma^{\mu} 
(1+\gamma^{5}) c \bar{c} \gamma_{\mu} (1+\gamma^{5} ) s | B^{0} 
\rangle , \\
\langle O^{LR} \rangle & = & \langle K_{S} \psi | \bar{b} \gamma^{\mu} 
(1+\gamma^{5}) c \bar{c} \gamma_{\mu} (1-\gamma^{5} ) s | B^{0} 
\rangle.
\end{eqnarray}
In the vacuum insertion approximation we find $\langle O^{LR} \rangle / 
\langle 
O^{RR} \rangle $ to be ${\cal O} (1)$.  With the upper bound (\ref{CLEOratio}) 
this gives
\begin{equation}
\frac{T^{LR}}{T^{RR}} \alt 3.55.
\end{equation}
Pollution is possibly over 100\% and this decay ceases to be predictive. (It 
is clean in the standard model due to CKM suppression of the penguins.)

In the same way we examine the decays $B_{d} \to D_{1}^{0}\pi^{0}, D^{+} 
D^{-}, 
K_{S} \pi^{0}, \phi K_{S}, K_{S} K_{S}$ and $B_{s} \to D_{S}^{+} D_{S}^{-}, 
D_{1}^{0} K_{S}, \psi K_{S}, \rho^{0} K_{S}, K^{+} K^{-}, \eta^{\prime} 
\eta^{\prime}, \phi K_{S}$.  For pure penguin decay processes we determine the 
mixing contribution by assuming that the ratio of left-right to right-right 
matrix elements is ${\cal{O}}(1)$.  For all of these decays we find pollution 
due to mixing on the order of 100\%.  There is little predictive power left 
from 
CP asymmetries in 
the neutral $B$ sector.  We stress that although disappointing this is a new 
result for models employing the GW ansatz.  This model is not inconsistent 
with any values of the various CP asymmetries in the neutral $B$ sector.  In 
fact, this model is consistent with no correlations of any kind among the 
phases in these decays.

\section{Conclusions}

In this $SU(2)_{L} \times SU(2)_{R} \times U(1)$ model where the third 
generation interacts weakly through 
pure right-handed couplings the parameters are highly constrained.   
Nevertheless, we have found 
a region in parameter space in which this model is consistent with 
measurements of the neutral meson mass differences $\Delta m_{K}$, $\Delta 
m_{B_{d}}$ and $\Delta m_{B_{s}}$ and semi-leptonic $B$ decays. We find that 
it 
is necessary that the ratio of coupling constants, $g_{R}/g_{L}$, be on the 
order of two.  
Constraining the phases with CP violating observables leads to a second 
undesirable result. There are three 
fine tuning conditions on the four phases in the right-handed mixing matrix. 

Constraints from CP violating observables also require that mixing between 
the left and right-handed $W$'s is not small, 
but $\zeta_{g} \sim {\cal{O}}(10^{-2})$. This leads to pollution in CP 
asymmetries in $B$ 
decays to CP eigenstates on the order of 100\%.  There are no definite 
predictions or clean phase measurements in these decays.  If discrepancies 
between experiment and the standard model are found in the $B$ decay 
asymmetries 
this model can not be ruled out.  However, increasing the lower bound on the 
right-handed $W$ from direct searches and a more stringent limit on 
$\zeta_{g}$ 
could decisively determine the fate of this model.

\acknowledgements
 
I thank Adam Falk for introducing the topic, helpful direction and many useful 
discussions.  This work 
was supported by the United States National Science Foundation under Grant 
No.~PHY-9404057 and by an Owen Fellowship of the Johns Hopkins University, 
Zanvyl Krieger School of Arts and Sciences.

\appendix

\section{Calculation of $\epsilon' / \epsilon$}

The expression (\ref{analeprime}) relates $\epsilon'$ to $A_{I}$, the neutral 
$K$ decay amplitude to pions with isospin $I$, given in (\ref{isoamp}).    
To calculate $\epsilon' / \epsilon$ we will use this expression employing an 
isoconjugate simplification 
due to 
Mohapatra and Pati~\cite{isoconj,Mohapatra:epsilon}.  
In this 
procedure the $\Delta S = 1$ weak decay Hamiltonian is decomposed into scalar 
($S$) and 
pseudoscalar terms ($P$) which are further decomposed into CP even 
and odd 
components (denoted by superscript $+$ or $-$).
\begin{equation}
 H^{\Delta S = 1} = S^{+} + S^{-} + P^{+} + P^{-}. 
 \end{equation}
If a relationship can be found such that
\begin{equation}
[I_{3} , P^{-} ] = i \alpha P^{+},
\end{equation}
where $\alpha$ is a real constant, then it can be shown that the ratio 
${\cal I}m A_{I} / {\cal R}e A_{I}$
is independent of $I$ and $\epsilon^{\prime} = 0$.  

To see this notice that with  $| K_{1,2} \rangle$ as CP eigenstates, $ I_{3} | 
K_{1} 
\rangle = -1/2 | K_{2} \rangle$ and $ I_{3} | \pi^{i} \pi^{j} \rangle = 0$ 
where $(i, j)$ denote $(+, -)$ or $(0, 0)$.   Now
\begin{equation}
\langle \pi^{i} \pi^{j} | P^{-} | K_{2} \rangle = i \alpha
\langle \pi^{i} \pi^{j} | P^{+} | K_{1} \rangle
\end{equation}
holds independent of $i$ and $j$.  The amplitude, $A_{I}$, can then be written 
as a real factor containing matrix elements multiplied by a complex factor 
independent of $I$.  The matrix elements cancel in ${\cal I}m A_{I} / {\cal 
R}e A_{I}$ and we have the desired result.  

In the GW model the $\Delta S = 1$ Hamiltonian can be split 
into three terms  
pertaining to the tree and penguin amplitudes for pure 
left-handed couplings, 
pure right-handed couplings and mixed couplings.  This can be 
written as
\begin{equation}
 H = (T^{LL} + P^{LL}) + (T^{RR} + P^{RR}) + (T^{LR} + P^{LR}),
\end{equation}
where $T$ and $P$ denote tree and penguin contributions 
respectively.  Because of 
the pure right-handed nature of the third generation couplings, of the penguin 
diagrams only $P^{RR}$ 
has a contribution from the top quark.  The elements have been 
calculated~\cite{Grimus:deltaS,Chia:dsgvert}.  
They are
\begin{eqnarray}
T^{LL} & = & \frac{4 G_{F}}{\sqrt{2}} (V_{ud}^{L} V_{us}^{L*}) 
\bar{s}_{L} 
\gamma^{\mu} u_{L} \bar{u}_{L} \gamma_{\mu} d_{L} + {\rm h.c.} ,
\\	
T^{RR} & = & \frac{4 G_{F}}{\sqrt{2}} \beta_{g} (V_{ud}^{R} 
V_{us}^{R*}) 
\bar{s}_{R} \gamma^{\mu} u_{R} \bar{u}_{R} \gamma_{\mu} d_{R} + 
{\rm h.c.} ,
\\ 
T^{LR} & = & \frac{4 G_{F}}{\sqrt{2}} \zeta_{g} \bigl( V_{ud}^{L} 
V_{us}^{R*} 
\bar{s}_{R} \gamma^{\mu} u_{R} \bar{u}_{L} \gamma_{\mu} d_{L} + V_{ud}^{R} 
V_{us}^{L*} \bar{s}_{L} \gamma^{\mu} u_{L} \bar{u}_{R} 
\gamma_{\mu} d_{R} \bigr) + 
{\rm h.c.},
\\
P^{LL} & = & \frac{4 G_{F}}{\sqrt{2}} \frac{\alpha_{s}(\mu)}{24 
\pi} \left( 
\sum_{q=u,c} V_{qd}^{L} V_{qs}^{L*} 
f \left( \frac{m_{q}^{2}}{M_{L}^{2}}\right) \right)  (\bar{u} \gamma_{\mu} 
\tau^{a} u + \bar{d} 
\gamma_{\mu} 
\tau^{a} d) 
\bar{s}_{L} \gamma^{\mu} \tau^{a} d_{L}  + {\rm h.c.},
\\
P^{RR} & = & \frac{4 G_{F}}{\sqrt{2}} \beta_{g} 
\frac{\alpha_{s}(\mu)}{24 \pi} 
\left( \sum_{q=u,c,t} V_{qd}^{R} V_{qs}^{R*} 
f \left( \frac{m_{q}^{2}}{M_{R}^{2}} \right) 
\right)  (\bar{u} \gamma_{\mu} \tau^{a} u + \bar{d} 
\gamma_{\mu} 
\tau^{a} d) 
\bar{s}_{R} \gamma^{\mu} \tau^{a} d_{R}  + {\rm h.c.} \\
P^{LR} & = &  \frac{4 G_{F}}{\sqrt{2}} \zeta_{g} 
\frac{\alpha_{s}(\mu)}{8 \pi}
(\bar{u} \gamma^{\mu} \tau^{a} u + \bar{d} \gamma^{\mu} \tau^{a} d) 
\frac{k^{\nu}}{k^{2}} \\
& & \qquad \times \bar{s} i \sigma_{\mu \nu} \left( 
\sum_{q=u,c} ( 
V_{qd}^{L} V_{qs}^{R*} \gamma_{L} + V_{qd}^{R} V_{qs}^{L*} 
\gamma_{R}) 
g\left( \frac{m_{q}^{2}}{M_{L}^{2}} \right) m_{q} \right) \tau^{a} d   + {\rm 
h.c.} ,\nonumber
\end{eqnarray}
where $f$ and $g$ are dimensionless functions of quark masses and 
$\gamma_{R/L}
= (1 \pm \gamma^{5})/2$.

These terms are now separated into scalar and pseudoscalar 
components.  In dealing 
with kaon decays to two pions only the pseudoscalar terms are 
relevant.  We want to decompose the Hamiltonian into a part which satisfies an 
isoconjugate relation, $H_{0}$, and a part which 
accounts for the nonzero $\epsilon^{\prime}$.  To do this we use the unitarity 
relations,
\begin{equation}
V_{ud}^{L} V_{us}^{L*} = - V_{cd}^{L} V_{cs}^{L*},
\end{equation}
\begin{equation}
V_{cd}^{R} V_{cs}^{R*} = - V_{ud}^{R} V_{us}^{R*} - V_{td}^{R} 
V_{ts}^{R*},
\end{equation}
to remove these two factors from the Hamiltonian.
The pseudoscalar part of the Hamiltonian which satisfies an isoconjugate 
relation can be written as
\begin{equation}
H_{0} = (- V_{ud}^{L} V_{us}^{L*} + \beta_{g} V_{ud}^{R} 
V_{us}^{R*} ) P + 
{\rm h.c.},
\end{equation}
where
\begin{eqnarray}
\lefteqn{P = \frac{G_{F}}{\sqrt{2}} \Biggl[ \bar{s} \gamma^{\mu} \gamma^{5} u 
\bar{u} 
\gamma_{\mu} d + 
\bar{s} \gamma^{\mu} u \bar{u} \gamma_{\mu} 
\gamma^{5} d} \nonumber \\ & & \hspace{1in} \mbox{} + 
\frac{\alpha_{s}(\mu)}{12 \pi} (\bar{u} \gamma_{\mu} \tau^{a} u 
+ \bar{d} 
\gamma^{\mu} \tau^{a} d) \bar{s} \gamma_{\mu} \gamma^{5} 
\tau^{a} d \left( 
f \left( \frac{m_{u}^{2}}{M_{L}^{2}} \right) - f \left( 
\frac{m_{c}^{2}}{M_{L}^{2}} \right) \right) \Biggr].
\end{eqnarray}
We have used the fact that the relation,
\begin{equation}
f \left( \frac{m_{u}^{2}}{M_{L}^{2}} \right) - f \left( 
\frac{m_{c}^{2}}{M_{L}^{2}} \right) =f \left( \frac{m_{u}^{2}}{M_{R}^{2}} 
\right) - f \left( 
\frac{m_{c}^{2}}{M_{R}^{2}} \right),
\end{equation}
holds up to negligible corrections of ${\cal{O}}(10^{-4})$.
Splitting this term and its Hermitian conjugate into CP even 
and odd states, 
$P^{+}$ and $P^{-}$ respectively, we arrive at the  
relationship
\begin{equation}
[ I_{3} , P^{-} ]  = -\frac{i {\cal I} m (- V_{ud}^{L} 
V_{us}^{L*} + \beta_{g} 
V_{ud}^{R} V_{us}^{R*} ) }{2 {\cal R} e (- V_{ud}^{L} 
V_{us}^{L*} + \beta_{g} 
V_{ud}^{R} V_{us}^{R*} ) } P^{+}.
\end{equation}

We now examine the mixing terms and the term 
proportional to $V_{td}^{R} V_{ts}^{R*}$.  We 
define the operators 
\begin{eqnarray}
O^{\pm} & = & \bar{s} \gamma^{\mu} \gamma^{5} u \bar{u} 
\gamma_{\mu} d \pm 
\bar{s} \gamma^{\mu} u \bar{u} \gamma_{\mu} \gamma^{5} d ,
\\
O^{5} & =  & \left( \bar{u} \gamma_{\mu} \tau^{a} u + \bar{d} 
\gamma_{\mu} \tau^{a} d \right) \bar{s} \gamma^{\mu} \gamma^{5} 
\tau^{a} d,
\\
O^{P}_{LR} & = & \frac{k^{\nu}}{k^{2}} \bar{s} i \sigma_{\mu \nu} \gamma^{5} 
\tau^{a} d  (\bar{u} \gamma^{\mu} \tau^{a} u + \bar{d} 
\gamma^{\mu} \tau^{a} 
d).
\end{eqnarray}
The dominant contributions to the ratios ${\cal I}m 
A_{I} / {\cal R}e 
A_{I}$ cancel in the difference of isospin channels.  In terms 
of the above 
operators the following ratios are needed.
\begin{eqnarray}
R^{P} & = & \frac{ \frac{\alpha_{s} (\mu)}{12 \pi} \left( f \left( 
\frac{m_{t}^{2}}{M_{R}^{2}} \right) - f \left( \frac{m_{c}^{2}}{M_{R}^{2}} 
\right) 
\right) \langle O^{5}_{1/2} \rangle }{ \langle O_{1/2}^{+} \rangle + 
\frac{\alpha_{s} (\mu)}{12 \pi} 
\left( f ( 
\frac{m_{u}^{2}}{M_{L}^{2}}) - f \left( \frac{m_{c}^{2}}{M_{L}^{2}} \right) 
\right) \langle O^{5}_{1/2} \rangle },
\\
R^{LR}_{u} & = & \frac{ \langle O_{1/2}^{-} \rangle - \frac{\alpha_{s} 
(\mu)}{4 
\pi} m_{q} g 
( 
\frac{m_{q}^{2}}{M_{L}^{2}}) \langle O^{P}_{LR 1/2} \rangle }{ \langle 
O_{1/2}^{+} \rangle 
+ 
\frac{\alpha_{s} (\mu)}{12 \pi} \left( f \left(
\frac{m_{u}^{2}}{M_{L}^{2}} \right) - f \left(
\frac{m_{c}^{2}}{M_{L}^{2}} \right) \right) \langle O^{5}_{1/2} 
\rangle}-\frac{\langle O_{3/2}^{-} \rangle}{\langle O_{3/2}^{+} \rangle},
\\
R^{LR}_{c} & = & \frac{ \frac{\alpha_{s} (\mu)}{4 \pi} m_{q} g 
( 
\frac{m_{q}^{2}}{M_{L}^{2}}) \langle O^{P}_{LR 1/2} \rangle }{ \langle 
O_{1/2}^{+} \rangle 
+ 
\frac{\alpha_{s} (\mu)}{12 \pi} \left( f \left(
\frac{m_{u}^{2}}{M_{L}^{2}} \right) - f \left(
\frac{m_{c}^{2}}{M_{L}^{2}} \right) \right) \langle O^{5}_{1/2} \rangle},
\end{eqnarray}
where we use $\langle O_{\Delta I} \rangle= \langle \pi \pi_{I} | O | K 
\rangle $.  To first order in $\zeta_{g}$ 
and $\beta_{g}$, $\epsilon^{\prime}$ is given by
\begin{eqnarray}
\epsilon^{\prime} & = &
\frac{\omega}{\sqrt{2}} \frac{1}{V_{ud}^{L} V_{us}^{L} } \Bigl(
\beta_{g}  
{\cal I } m (V_{td}^{R} V_{ts}^{R*} )  R^{P}   \nonumber \\
& & \mbox{} + \zeta_{g} {\cal 
I} m 
(V_{ud}^{L} V_{us}^{R*} - V_{ud}^{R} V_{us}^{L*} ) R^{LR}_{u}  + \zeta_{g} 
{\cal 
I} m (V_{cd}^{L} V_{cs}^{R*} - 
V_{ud}^{R} V_{us}^{L*} ) 
R^{LR}_{c} \Bigr),
\end{eqnarray}
where $\omega = {\cal{R}}e A_{2}/{\cal{R}}e A_{0} \approx 1/20$.

There are hadronic uncertainties in the ratios of matrix elements.  To 
restrict 
the parameters in this model we need to obtain order of magnitude 
estimates of these ratios at the least. Assuming an ${\cal O} (1)$ estimate 
for 
$\langle O^{5} \rangle / 
\langle O^{+} \rangle$ and $ \langle O^{-} \rangle / \langle O^{+} \rangle$ we 
obtain $R^{P} \sim {\cal O} (10^{-1})$.    
Assuming 
$k^{\nu}/k^{2} \sim 1 \; {\rm GeV}^{-1}$ we estimate $R_{c}^{LR} \sim {\cal O} 
(10^{-1})$ and $R^{LR}_{u} \sim {\cal O} (1)$.
 
Evaluating the coefficients of the operators after imposing  
(\ref{phase1}), the constraint from $\epsilon$, we find the following 
expression. 
\begin{equation}
\epsilon^{\prime}/ \epsilon = \zeta_{g} \left( 1.8 R_{c}^{LR} \sin 
(\alpha-\beta) 
-4.0 R_{u}^{LR} \sin(\alpha + \beta) \right)
\end{equation}
We use this result in section \ref{CPconstraint}.

\end{document}